\documentstyle[12pt]{article}
\def\al{\alpha}

\def\ga{\gamma}
\def\de{\delta}
\def\ep{\epsilon}

\def\et{\eta}

\def\si{\sigma}

\def\ps{\psi}

\def\Ga{\Gamma}
\def\De{\Delta}

\def\fr#1#2{{{#1} \over {#2}}}

\def\bra#1{\langle{#1}|}
\def\ket#1{|{#1}\rangle}

\def\expect#1{\langle{#1}\rangle}

\def\half{{\textstyle{1\over 2}}}
\def\frac#1#2{{\textstyle{{#1}\over {#2}}}}

\def\lsim{\mathrel{\rlap{\lower4pt\hbox{\hskip1pt$\sim$}}
    \raise1pt\hbox{$<$}}}
\def\gsim{\mathrel{\rlap{\lower4pt\hbox{\hskip1pt$\sim$}}
    \raise1pt\hbox{$>$}}}
\def\sqr#1#2{{\vcenter{\vbox{\hrule height.#2pt
         \hbox{\vrule width.#2pt height#1pt \kern#1pt
         \vrule width.#2pt}
         \hrule height.#2pt}}}}

\def\Re{\hbox{Re}\,}
\def\Im{\hbox{Im}\,}

\newcommand{\beq}{\begin{equation}}
\newcommand{\eeq}{\end{equation}}
\newcommand{\bea}{\begin{eqnarray}}
\newcommand{\eea}{\end{eqnarray}}
\newcommand{\rf}[1]{(\ref{#1})}
\newcommand{\ct}[1]{\cite{#1}}
\newcommand{\eq}[1]{Eq.\ \rf{#1}}

\renewenvironment{thebibliography}[1]
 { \rm
   \begin{list}{\arabic{enumi}.}
    {\usecounter{enumi} \setlength{\parsep}{0pt}
     \setlength{\itemsep}{3pt} \settowidth{\labelwidth}{#1.}
     \sloppy
    }}{\end{list}}

\begin{document}
\titlepage

\begin{flushright}
{IUHET 307\\}
{hep-ph/9510365\\}
{June 1995\\}
\end{flushright}
\vglue 1cm

\begin{center}
{{\bf TESTING CPT WITH THE NEUTRAL--D SYSTEM
\\}
\vglue 1.0cm
{Don Colladay and V. Alan Kosteleck\'y\\}
\bigskip
{\it Physics Department\\}
\medskip
{\it Indiana University\\}
\medskip
{\it Bloomington, IN 47405, U.S.A.\\}

\vglue 0.8cm
}
\vglue 0.3cm

\end{center}

{\rightskip=3pc\leftskip=3pc\noindent
We investigate the issue of testing CPT invariance
in the neutral-$D$ system,
using $D$ events obtained either from fixed-target experiments
or from a tau-charm factory.
For both types of experiments
we show that the expected suppression of mixing in the $D$ system,
normally viewed as a disadvantage for CP tests,
allows unsuppressed measurement of certain parameters
describing CPT violation.
Asymmetries are presented that permit
the extraction of parameters
for direct CPT violation in the $D$ system
and for indirect CPT violation in the $K$ system.
We also show that experiments on the neutral-$D$ system
provide an alternative means for measuring
conventional indirect $T$ violation in the kaon system.

}

\vskip 1truein
\centerline{\it Accepted in Rapid Communications,
Physical Review D, issue of December 1995}

\vfill
\newpage

\baselineskip=20pt

{\bf\noindent I. Introduction}
\vglue 0.4cm

CPT invariance is believed to be a fundamental symmetry
of local relativistic point-particle field theories
\cite{s1,s2,s3,s4}.
Experimental studies of CPT symmetry
probe the foundations of modern particle physics,
and they can therefore provide tests
of certain alternatives to conventional theories.
For example,
one measurable signature of an underlying string theory
could be a violation of CPT appearing
through a mechanism ultimately traceable to
string nonlocality
\cite{kp1,kp4}.
Another example is a possible CPT-violating effect
that might arise from modifications of quantum mechanics
due to quantum gravity
\cite{sh,dp,rw},
perhaps in the context of string theory
\cite{emn}.

The most stringent bound on CPT violation,
which by one measure is a few parts in $10^{18}$
\cite{c1,c2,c3},
arises from interferometric tests in the neutral-kaon system.
Other neutral-meson systems
may also provide interesting CPT bounds.
For example,
in the string-based scenario for CPT violation,
effects could appear at levels accessible to experiment
not only in the $K$ system but also in the $B$ and $D$ systems
\cite{kp4}.
Since the corresponding sizes of any CPT violation
could be different,
it is important to bound experimentally the
CPT-violating parameters in all these systems.

To date,
no experimental bounds have been placed
on CPT violation in the $B$ or $D$ systems.
The feasibility of placing experimental limits
on the parameters for direct and indirect CPT violation
in the $B$ system
has recently been demonstrated
\ct{ck}.
It is plausible that analysis of existing data could already set
interesting bounds,
and the $B$ factories currently under construction
should provide further improvements.

In contrast,
despite significant advances in the
understanding of charm physics
(for a recent proceedings, see ref.\ \ct{c2000}),
there has been as yet neither an experimental study
nor a systematic theoretical treatment
of CPT violation in the $D$ system.
Indeed,
measurements of \it any \rm
indirect CP violation in the $D$ system
are generally viewed as infeasible.
This is
because the decay time for the neutral $D$ meson
is much less than its characteristic mixing time,
so any indirect CP violation is suppressed
by a small mixing parameter $x$.
Also,
the short $D$ lifetime means that only time-integrated
rates can be observed.
This raises a second issue:
disentangling the various T and CPT effects
so the quantities parametrizing CPT violation
can be isolated.

In this paper,
we address the issues of suppression and disentanglement
with a model-independent treatment
in the context of conventional quantum mechanics.
Our framework can therefore handle the string-based CPT violation
discussed in refs.\ \cite{kp1,kp4}.
Additional CPT-violating effects involving
some modified form of quantum mechanics
might arise from the evolution of pure states into mixed ones
in quantum gravity
\cite{sh,dp,rw}
or possibly in string theory
\cite{emn}.
One approach to modeling such effects
is to modify the Schr\"odinger equation,
which introduces additional parameters.
In the context of the kaon system,
a simple parametrization
involving three additional quantities has been suggested
\cite{ehns,rp}.
A treatment of this topic in the present context
lies outside the scope of the present paper.
We restrict ourselves here to the observation
that it would be of interest to generalize the present results
and those of ref.\ \cite{ck}
to incorporate possible CPT violation in the $B$ and $D$ systems
arising from modifications to quantum mechanics.

To address the issue of suppression,
we show that the size of the mixing parameter $x$
is irrelevant to the measurement of certain parameters
describing direct CPT and indirect T violation.
Moreover,
the suppression of indirect $D$-system CP violation
can in fact be viewed as an advantage
because the $D$-decay modes to kaons
are then accompanied by unsuppressed
and measurable indirect $K$-system CPT and T violation.
These results indicate that CP studies
in the neutral-$D$ system are of experimental interest
despite the small size of the mixing parameter $x$.

To address the issue of disentangling T and CPT effects,
we provide certain asymmetries that separate parameters
describing CPT violation.
In the analysis,
we consider fixed-target experiments producing
single tagged neutral $D$ mesons
(these would also include tagged $D$ mesons from
a $B$ factory)
and experiments at a $\tau$-charm factory,
which would produce large numbers of correlated
$D^0 \overline{D^0}$ pairs from the decay of the
$\psi(3770)$ resonance.
We obtain estimates
for bounds on CPT violation that could be obtained
from present and future experiments of both classes.

Current experimental limits on direct CP violation
in CKM-suppressed decay modes of the $D$ meson
are attaining the 10\% level
\cite{pds}.
About $10^5$ fully reconstructed charm events already exist
\cite{rjm},
and it appears feasible to obtain about
$10^8$ fully reconstructed $D$ events
by the turn of the century
using fixed-target and factory experiments
\cite{jaa}.
The analysis we present here suggests that
some bounds on CPT violation
might already be obtained from extant data
and that results of experiments over the next
few years would yield useful limits.

\vglue 0.6cm
{\bf\noindent II. Preliminaries}
\vglue 0.4cm

The eigenvectors of the effective hamiltonian
for the $D^0$-$\overline{D^0}$ system
are\footnote{Throughout this paper
we assume small CP violation,
implying small T and CPT violation,
and we neglect terms that are higher-order
in small quantities.
Our phase conventions are discussed in more detail
in refs.\ \ct{kp4,ck}. }
\bea
\ket{D_S} & = & \frac 1 {\sqrt 2}
[(1 + \ep_D + \de_D)\ket{D^0}
+(1 - \ep_D - \de_D)\ket{\overline{D^0}}]
\quad , \nonumber\\
\ket{D_L} & = & \frac 1 {\sqrt 2}
[(1 + \ep_D - \de_D)\ket{D^0}
-(1 - \ep_D + \de_D)\ket{\overline{D^0}}]
\quad .
\label{iia}
\eea
The CP-violating complex parameters
$\ep_D$ and $\de_D$ are measures
of indirect T and indirect CPT violation, respectively.
The analogous parameters
$\ep_K$ and $\de_K$ for the $K$ system
are defined by Eq.\ \rf{iia}
but with the replacement $D\to K$.

We denote the decay rates
of the physical particles $D_S$, $D_L$
by $\ga_S$, $\ga_L$
and their masses by $m_S$, $m_L$.
Useful combinations
of these basic parameters are
$\De \ga = \ga_S - \ga_L$,
$\De m = m_L - m_S$,
$\ga=\ga_{S}+\ga_{L}$,
$a^2 = \De m^2 + \De \ga^2/4$,
and $b^2 = \De m^2 + \ga^2/4$.

In the $D$ system,
the mixing parameter $x = 2\De m / \ga$
is experimentally
bounded\footnote{The analysis leading to this result
assumes negligible CPT violation.
The value of $x$ could be larger if
significant CPT violation is present.
Our assumption of small CP violation
makes it consistent to take $x$ small also.}
\cite{pdt}
to $|x| < 0.08$.
The theoretical value of $x$ is uncertain.
Calculations based on standard-model physics
\cite{hg}
suggest that $|x|$ is smaller than $10^{-2}$,
although the presence of long-distance dispersive effects
makes accurate prediction difficult
\cite{wolf,dght}.
However,
extensions to the standard model can
generate larger values of $x$.
We therefore keep terms to order $x^2$
in what follows.
Also,
in explicit estimates,
we take $\De \ga \simeq \De m$ for simplicity.
Among the effects of scaling $\De \ga$ relative to $\De m$
is the scaling of $\Re \de_D$ relative to $\Im \de_D$.
In any event,
any such effects are straightforward to calculate
from the general expressions we provide below.

Our analysis makes use of
two different classes of $D$-meson decays.
The first class,
called {\it semileptonic-type} $f$ {\it decays},
includes the usual semileptonic decays along with a
special class of other modes $D^0 \rightarrow f$ for which
there is no lowest-order weak process
allowing a significant contamination of either
$\overline {D^0} \rightarrow f$
or $D^0 \rightarrow \overline f$.
The currently observed final states $f$ of
this type with significant branching ratios
are the usual semileptonic ones
and those that involve production of
a $\overline K^*(892)^0$ along
with other non-strange mesons.
For the remaining observed modes,
a CKM-suppressed process contributes
to the contaminating transitions.
The second class,
called {\it semileptonic-type}
$\overline{K^0} X$ {\it decays},
includes all final states that contain a
$\overline{K^0}$ but do {\it not} contain a $K^0$.
This means there is no lowest-order weak process
for $\overline {D^0} \rightarrow \overline{K^0} X$
or $D^0 \rightarrow K^0 \overline{X}$.
Note that the above requirements for both types of decay
are more stringent than imposing only
that the conjugate mode is CKM suppressed.
For example,
the state $f \equiv K^-\pi^+$ is excluded.

\vglue 0.6cm
{\bf\noindent III. Fixed-Target Experiments}
\vglue 0.4cm

This section presents our analysis of
rates and asymmetries for fixed-target experiments.
We first consider
integrated decay rates involving
semileptonic-type $f$ decays.
The associated transition amplitudes
can be parametrized as
\cite{lw,td}:
\bea
\bra{f}T\ket{D^0} = F_f (1 - y_f)~~~~,~~ &
\bra{f}T\ket{\overline{D^0}} = x_f F_f (1 - y_f)
\quad , \nonumber \\
\bra {\overline f}T\ket {\overline{D^0}} =
F_f^*(1 + y_f^*)~~~~,~~ &
\bra {\overline f}T\ket{D^0} = {\overline x_f^*} F_f^* (1 + y_f^*)
\quad .
\label{iiia}
\eea
The independent quantities $F_f$, $y_f$, $x_f$ and
$\overline x_f$ are all complex.
The latter two vanish if $\De C = \De Q$,
so in what follows we treat them as small.
If T invariance holds,
all four quantities are real.
If CPT invariance holds,
$x_f = \overline x_f$ and $y_f = 0$.
The parameter $y_f$ therefore characterizes
direct CPT violation in the decay to $f$
and as such is of particular interest here.

Time-integrated rates
for semileptonic-type $f$ decays of $D$ mesons
can be expressed in terms of the above quantities.
Denote by $\ket{D(t)}$ the time-evolved state arising from
the state $\ket{D^0}$ at $t=0$
and by $\ket{\overline{D}(t)}$ the state
arising from $\ket{\overline{D^0}}$ at $t=0$.
Then,
there are four time-integrated rates of interest,
given by
\bea
R_f & = & \int_{0}^{\infty} dt |\bra{f}T\ket{D(t)}|^2
\nonumber \\
& = & |\fr{F_f} {2}|^2 \Bigl[ \ga (\fr{1}{\ga_S \ga_L}
+\fr{1}{b^2})(1 - 2\Re y_f) - \fr{2 \De \ga}{\ga_S \ga_L}
\Re (2\de_D + x_f) \nonumber \\
& & - \fr {4 \De m}{b^2} \Im (2\de_D + x_f)
\Bigr]
\label{rf1}
\quad ,
\eea
\beq
\overline{R}_{\overline f} = \int_{0}^{\infty} dt
|\bra{\overline f}T\ket{\overline D(t)}|^2
= R_f(y_f \rightarrow -y_f, \de_D \rightarrow -\de_D,
x_f \rightarrow \overline{x_f}^*)
\label{rf2}
\quad ,
\eeq
\bea
R_{\overline f} & = & \int_{0}^{\infty} dt
|\bra{\overline f}T\ket{D(t)}|^2
\nonumber \\
& = & |\fr{F_f} {2}|^2 \Bigl[ \ga (\fr{1}{\ga_S \ga_L}
-\fr{1}{b^2})(1 - 2\Re (2\ep_D - y_f))
- \fr{2 \De \ga}{\ga_S \ga_L} \Re \overline x_f \nonumber \\
& & - \fr {4 \De m}{b^2} \Im \overline x_f \Bigr]
\label{rf3}
\quad ,
\eea
\beq
\overline{R}_{f} = \int_{0}^{\infty} dt
|\bra{f}T\ket{\overline D(t)}|^2
= R_{\overline f}(y_f \rightarrow -y_f,
\ep_D \rightarrow -\ep_D,\overline{x_f} \rightarrow x_f^*)
\label{rf4}
\quad .
\eeq
The expressions for the rates $\overline{R}_{\overline f}$ and
$\overline{R}_f$ are obtained by making the indicated
replacements in $R_f$ and $R_{\overline f}$,
respectively.

{}From these rates,
we can extract an asymmetry providing information
about direct CPT violation.
It is
\bea
A_f & \equiv & \fr{R_f - \overline R_{\overline f}}
{R_f + \overline R_{\overline f}}
\nonumber \\
& = & -2 \Re y_f - \fr{1}{\ga (b^2 + \ga_S \ga_L)}
\Bigl[\De \ga b^2(4 \Re \de_D +\Re (x_f - \overline x_f))
\nonumber \\
& & + 2 \De m \ga_S \ga_L (4 \Im \de_D + \Im (x_f +
\overline x_f))\Bigr]
\nonumber \\
& \simeq & -2 \Re y_f -x[\Re \de_D + \fr 1 {4}
\Re (x_f - \overline x_f) + 2 \Im \de_D + \fr 1 {2}
\Im (x_f + \overline x_f)]
\quad .
\label{if}
\eea
The first term in this expression
is a measure of direct CPT violation in the $D$ system.
The remaining terms are suppressed by
the mixing parameter $x$.
If we further assume that violations of
$\De C = \De Q$ are independent of CP violation,
then $x_f = \overline x_f^*$.
This makes the asymmetry $A_f$ independent of
the particular final state.

Another asymmetry can be formed as follows:
\beq
A_f^{\prime} \equiv \fr{\overline R_f - R_{\overline f}}
{\overline R_f + R_{\overline f}}
\approx 2\Re(2\ep_D - y_f)
\quad .
\eeq
In deriving the latter expression,
we have taken violations of $\De C = \De Q$
to vanish for simplicity.
This asymmetry is of lesser interest
than that of \eq{if} because
the rates themselves are suppressed by at least
one power of $x$.
This means the statistics required
for observation of this asymmetry
are increased by a factor of at least $1/x^2$,
so we disregard it in the following.

Next,
we consider integrated decay rates involving
semileptonic-type $\overline{K^0} X$ decays.
Define transition amplitudes for these processes
analogous to the definitions in Eq.\ \rf{iiia}:
\bea
\bra{\overline K^0}T\ket{D^0} = F_K (1 - y_K)~~~~,~~ &
\bra{\overline K^0}T\ket{\overline{D^0}} = x_K F_K (1 - y_K)
\quad , \nonumber \\
\bra {K^0}T\ket {\overline{D^0}} =
F_K^*(1 + y_K^*)~~~~,~~ &
\bra {K^0}T\ket{D^0} = {\overline x_K^*} F_K^* (1 + y_K^*)
\quad .
\label{iiiia}
\eea
To simplify notation,
in the above amplitudes the symbols $\bra{K^0}$
and $\bra{\overline K^0}$ are used to represent the full
multiparticle final states
containing the corresponding $K$ meson.
The complex parameters $x_K$, $\overline x_K$, $F_K$, and $y_K$
depend on the specific final state involved.
They have the same properties under T and CPT invariance
as the analogous parameters
for semileptonic-type $f$ decays.
Note that $x_K$ and $\overline x_K$ are
taken to be small
because the associated amplitudes have no lowest-order
weak contribution.

Observable final states involve $K_S$ and $K_L$
rather than $K^0$ and $\overline K^0$.
We use the notation $\bra{K_S}$ and
$\bra{K_L}$ to represent the linear
combinations of the semileptonic-type
$\overline{K^0} X$ final state and its charge conjugate
given by the kaon equivalent of \eq{iia}.
With these amplitudes,
there are again four integrated rates:
\bea
R_S & = & \int_{0}^{\infty} dt |\bra{K_S}T\ket{D(t)}|^2
\nonumber \\
& = & \fr{\Re ^2 F_K}{2} \Bigl[\fr{1} {\ga_S}
+ (\fr{2}{\ga_S} - \fr{\ga}{b^2})
\Re (\de_D - \ep_D + \fr{1} {2} x_K)
+ (\fr{2}{\ga_S} + \fr {\ga}{b^2}) \Re \fr{1}{2}
\overline x_K
\nonumber \\
& & - \fr{\ga}{b^2} \Re (\ep_K + \de_K + y_K)
+ \fr{2 \De m}{b^2} (\Im (\ep_K + \de_K - \de_D + \ep_D
- \fr{1}{2}(\overline x_K + x_K))
\nonumber \\
& & + \fr{\Im F_K}{\Re F_K}) \Bigr]
\label{rs1}
\quad , \\
\overline R_S & = & \int_{0}^{\infty} dt
|\bra{K_S}T\ket{\overline D(t)}|^2
\nonumber \\
& = & R_S(\ep_D \rightarrow -\ep_D^*, \de_D \rightarrow
-\de_D^*, y_K \rightarrow -y_K, x_K \leftrightarrow
\overline{x}_K, \ep_K \rightarrow -\ep_K^*,
\de_K \rightarrow -\de_K^*)
\nonumber
\quad ,\\
&& \label{rs2}
\\
R_L & = & \int_{0}^{\infty} dt |\bra{K_L}T\ket{D(t)}|^2
\nonumber \\
& = & \fr{\Re ^2 F_K}{2} \Bigl[\fr{1} {\ga_L}
- (\fr{2}{\ga_L} - \fr{\ga}{b^2})
\Re (\de_D + \ep_D + \fr{1} {2} x_K)
- (\fr{2}{\ga_L} + \fr {\ga}{b^2}) \Re \fr{1}{2}
\overline x_K
\nonumber \\
& & - \fr{\ga}{b^2} \Re (\ep_K - \de_K + y_K)
+ \fr{2 \De m}{b^2} (\Im (\ep_K - \de_K + \de_D + \ep_D
+ \fr{1}{2}(\overline x_K + x_K))
\nonumber \\
& & + \fr{\Im F_K}{\Re F_K}) \Bigr]
\label{rl1}
\quad , \\
\overline R_L & = & \int_{0}^{\infty} dt
|\bra{K_L}T\ket{\overline D(t)}|^2
\nonumber \\
& = & R_L(\ep_D \rightarrow -\ep_D^*, \de_D \rightarrow
-\de_D^*, y_K \rightarrow -y_K, x_K \leftrightarrow
\overline{x}_K, \ep_K \rightarrow -\ep_K^*,
\de_K \rightarrow -\de_K^*)
\quad .
\nonumber\\
\label{rl2}
\eea
In the above a double arrow $\leftrightarrow$ is used to
indicate an interchange of parameters rather than a
substitution.

{}From these integrated rates,
two useful asymmetries can be constructed:
\bea
A_S & \equiv & \fr {\overline R_S - R_S}
{\overline R_S + R_S}
\nonumber \\
& = & 2(1 - \fr{\ga \ga_S}{2 b^2}) \Re (\ep_D - \de_D)
- \fr{\ga \ga_S}{2 b^2} \Re (\overline x_K - x_K)
+ \fr{\ga \ga_S}{b^2} \Re (\ep_K + \de_K + y_K)
\nonumber \\
& \simeq &
2(1 + \half x - x^2) \Re (\ep_K + \de_K + y_K
-\half (\overline{x}_K - x_K))
\nonumber \\
&  &
\qquad \qquad \qquad \qquad \qquad \qquad \qquad
-x(1 - 2x) \Re (\ep_D - \de_D)
\quad ,
\label{iif}
\\
A_L & \equiv & \fr {\overline R_L - R_L}
{\overline R_L + R_L}
\nonumber \\
& = & 2(1 - \fr{\ga \ga_L}{2 b^2}) \Re (\ep_D + \de_D)
+ \fr{\ga \ga_L}{2 b^2} \Re (\overline x_K - x_K)
+ \fr{\ga \ga_L}{b^2} \Re (\ep_K - \de_K + y_K)
\nonumber \\
& \simeq &
2 (1 - \half x - x^2) \Re (\ep_K - \de_K + y_K
+\half(\overline{x}_K - x_K))
\nonumber \\
& &
\qquad \qquad \qquad \qquad \qquad \qquad \qquad
+x(1 + 2x)\Re (\ep_D + \de_D)
\quad .
\label{iiif}
\eea
Their difference gives
the combination
\bea
A_L - A_S & = &
- 4 \Re \de_K
+ 2\Re (\overline x_K - x_K)
\nonumber \\
&+& 2 x \Re (\ep_D - \ep_K - y_K)
+ 4 x^2 \Re (\de_D + \de_K - \half (\overline{x}_K - x_K))
\quad ,
\label{ig}
\eea
while their sum is
\bea
A_L + A_S & = &
4\Re (\ep_K + y_K)
\nonumber \\
&+& 2x \Re (\de_D + \de_K - \half (\overline{x}_K -x_K))
+ 4x^2 \Re (\ep_D - \ep_K - y_K)
\quad .
\label{iig}
\eea
Assuming violations of $\De C = \De Q$
are independent of CPT violation,
the terms containing $\overline{x}_K - x_K$ vanish.
In any event,
for negligible mixing $x$ these expressions reduce to
their first terms.

\vglue 0.6cm
{\bf\noindent IV. Experiments at a Tau-Charm Factory}
\vglue 0.4cm

In this section,
we turn to a discussion of rates and asymmetries
for experiments that are feasible at a tau-charm factory.
Several relevant integrated rates arise.
One is the time-integrated rate $\Ga(f_1,f_2)$
for the decay of the correlated $D^0 \overline{D^0}$ pair
into states $f_1$ and $f_2$.
In discussing other rates,
it is convenient to separate the rate $\Ga(f_1,f_2)$
into the component $\Ga^+(f_1,f_2)$
for which the decay into $f_1$ occurs first
and the component $\Ga^-(f_1,f_2)$ for which $f_2$
occurs first.
Then,
another useful quantity
is the inclusive rate $\Ga^+_{\rm incl}(f_1)$,
obtained by summing $\Ga^+(f_1,f_2)$
over final states $f_2$.
We find
\bea
\Ga(f_1,f_2) & = & \fr 1 {2\ga_S\ga_L}
\Bigl[ |a_{1S}a_{2L}|^2 + |a_{1L}a_{2S}|^2 - \fr{\ga_S\ga_L}{b^2}
(a^*_{1S}a^*_{2L}a_{1L}a_{2S} + {\rm c.c.})\Bigr]
\quad ,
\label{rate} \\
\Ga^+_{\rm incl}(f_1) & = & \sum_{f_2}\Ga^+(f_1,f_2)
\nonumber \\
& = & \fr 1{2\ga} \Bigl[ |a_{1S}|^2 + |a_{1L}|^2
- 2[a^*_{1S} a_{1L}(\Re \ep_D + i \Im \de_D) + {\rm c.c.}]
\Bigr]
\quad ,
\label{incl}
\eea
where the transition amplitudes are defined as
\beq
a_{\al S} = \bra{f_\al}T\ket{D_S}~~,~~~~
a_{\al L} = \bra{f_\al}T\ket{D_L}~~,
\quad
\label{iid}
\eeq
Further details about calculating these rates
may be found in
refs.\ \ct{kp4,ck}.

We begin by considering semileptonic-type $f$ decays.
The parameter $\Re y_f$ describing direct CPT violation
can be extracted from the above rates
using the asymmetry
\beq
A^+_f \equiv \fr
{\Ga^+_{\rm incl}(f) - \Ga^+_{\rm incl}(\overline f)}
{\Ga^+_{\rm incl}(f) + \Ga^+_{\rm incl}(\overline f)}
= -2 \Re y_f
\quad .
\label{a}
\eeq
We see that this asymmetry provides a test
of direct CPT violation
that is independent of any T or indirect CPT violation.

Other asymmetries with suppressed component rates
or suppressed magnitudes can be examined
by restricting attention to semileptonic-type $f$
decays in both channels.
An asymmetry isolating parameters for indirect CPT violation
in the $D$ system can be constructed as
\bea
A_{f,\overline f} & \equiv & \fr
{\Ga^+(f,\overline f) - \Ga^-(f,\overline f)}
{\Ga^+(f,\overline f) + \Ga^-(f,\overline f)}
\nonumber \\
& = & 4 \fr {b^2 \De \ga \Re \de_D +
2 \De m \ga_S \ga_L \Im \de_D}
{\ga (b^2 + \ga_S \ga_L)}
\nonumber \\
& \simeq & x(\Re \de_D + 2 \Im \de_D)
\quad .
\label{jk}
\eea
The derivation assumes
violations of $\De C = \De Q$ are independent of
CP violation,
so that $x_f = \overline x_f^*$.
The $x$-dependence of the result means
that the number of events
required to measure a non-zero value for
$A_{f,\overline f}$ is scaled by $1/x^2$.
Another possible asymmetry is
\beq
A^{tot}_{f,\overline f} \equiv \fr
{\Ga(f,f) - \Ga(\overline f, \overline f)}
{\Ga(f,f) + \Ga(\overline f, \overline f)}
= 4 \Re (\ep_D - y_f)
\quad .
\eeq
In this case,
the rates themselves are suppressed
by a factor of $x^2$,
so the required statistics
are scaled by $1/x^4$.

Next,
we consider the case with one channel
involving a semileptonic-type $K_S X$ or $K_L X$ decay
while the other channel
involves a semileptonic-type $f$ decay.
Transition amplitudes for the $\overline K X$ decays are
defined in Eq.\ \rf{iiiia}.
With these,
we obtain:
\bea
\et_{K_S} & \equiv &
\fr{\bra{K_S}T\ket{D_L }}
{\bra{K_S}T\ket{D_S}}
\nonumber \\
& = &
- \ep_K^* + \ep_D - \de_K^* - \de_D
- \fr {\Re (F_{K} y_{K})} {\Re F_{K}}
- \fr 1 {2\Re F_{K}}
(x_{K} F_{K} - \overline x^*_{K} F^*_{K})
\nonumber \\
& & +~ i \fr {\Im F_{K}} {\Re F_{K}}
\Bigl[ 1
- i \fr {\Im F_{K}} {\Re F_{K}}
(- \ep_K^* + \ep_D - \de_K^* + \de_D )
\nonumber \\
& & \qquad\qquad
- \fr 1 {2\Re F_{K}}
(x_{K} F_{K} + \overline x^*_{K} F^*_{K})
+ i \fr {\Im (F_{K} y_{K})} {\Re F_{K}}
\Bigr]
\quad ,
\eea
\beq
\et_{K_L} \equiv
\fr{\bra{K_L}T\ket{D_S}}
{\bra{K_L}T\ket{D_L}} =
\et_{K_S}(\de_D \rightarrow -\de_D, \de_K \rightarrow
-\de_K, x_K \rightarrow -x_K, \overline{x}_K
\rightarrow -\overline{x}_K)
\quad .
\eeq

Using these,
we calculate two useful rate asymmetries:
\bea
A_{f,K_S} & \equiv &
\fr{\Ga(f,K_S) - \Ga(\overline f,K_S)}
{\Ga(f,K_S) + \Ga(\overline f,K_S)}
\nonumber \\
& = & 2\Re (\ep_D - y_f - \de_D)
- \fr{2\ga_S\ga_L}{b^2}\Re (\et_{K_S})
\quad ,
\label{ivc}
\eea
and
\beq
A_{f,K_L}  \equiv
A_{f,K_S} (K_S \to K_L) =
A_{f,K_S} (\de_D \to -\de_D,
\et_{K_S} \to \et_{K_L})
\quad .
\label{ivd}
\eeq
These two asymmetries have been derived under the
assumption that violations of $\De C=\De Q$
are independent of CPT violation,
so that
$x_f = \overline x_f$ and
$x_{K} = \overline x_{K}$.
Since $\Im F_{K}$ controls
the direct T violation in these processes,
we have also treated it as a small quantity.
This latter assumption is made more plausible
by the current experimental bounds
at about the $10 \%$ level
on parameters describing direct T violation
\cite{pds}.
The above expressions can be approximated
for the $D$ system as
\beq
A_{f,K_S} \simeq
- 2\Re y_f + 2\Re (\ep_K + \de_K + y_K)
+\frac 52 x^2 \Re (\ep_D - \de_D)
\label{vc}
\eeq
\beq
A_{f,K_L} \simeq
- 2\Re y_f + 2\Re (\ep_K - \de_K + y_K)
+\frac 52 x^2 \Re (\ep_D + \de_D)
\label{vd}
\eeq

The difference between the two above equations
is a function of CPT-violating parameters:\footnote{It
can be shown that Eq.\ \rf{ive} is in fact correct to terms
simultaneously quadratic in
$\Im F_{K}$ and linear in $\de_D$ or $\de_K$
even if the constraint
of small direct T violation is relaxed.}
\beq
A^-_{L,S} \equiv
A_{f,K_L} - A_{f,K_S}
\simeq - 4\Re \de_K + 5x^2 \Re \de_D
\quad .
\label{ive}
\eeq
The coefficient of $\Re \de_D$ is
of order no larger than $10^{-4}$.
The second term can therefore be neglected,
and we are left with an asymmetry measuring
the parameter $\Re \de_K$
for indirect CPT violation in the $K$ system.
This result is independent of either of the final states,
so the statistics can be made more favorable by
summing over the class of relevant final states.

The sum of the two asymmetries gives the combination
\beq
A^+_{L,S} \equiv
A_{f,K_L} + A_{f,K_S}
\simeq
4\Re (\ep_K + y_K - y_f)
+ 5x^2 \Re \ep_D
\quad .
\label{ve}
\eeq
Here,
the parameter measuring indirect T violation
in the neutral-$D$ system is suppressed by the factor
$x^2$,
thereby producing an asymmetry measuring the combination
$\Re (\ep_K + y_K - y_f)$.
Note that this quantity depends on both final states
through the direct CPT-violation parameters
$y_K$ and $y_f$.

\vglue 0.6cm
{\bf\noindent V. Estimates of Bounds Attainable}
\vglue 0.4cm

The asymmetries given in the previous two sections
demonstrate that CPT information
can be extracted from the $D$ system.
Next,
we investigate the bounds attainable
using these asymmetries.
For fixed-target experiments,
we estimate the number of $D^0$ particles
required to reduce the error in
a given asymmetry to one standard deviation.
For measurements at a tau-charm factory,
we estimate the number of $\ps(3770)$ events
required for a similar precision.
The analysis follows the methods presented in
refs.\ \cite{dhr,kp4,ck}.

Assuming a binomial event distribution
and a general asymmetry
$A = (N_+ - N_-)/(N_+ + N_-)$,
observation of a nonzero $\expect{A}$
at the $N\si$ level
requires an expected number of events
$\expect{N_+}
=N^2(1 + \expect{A})(1 - \expect{A}^2)/2\expect{A}^2$.
Conversion of this to an expected number of $D^0$ events
required in a fixed-target experiment
involves multiplication by the inverse branching ratio
for the $D^0$ decay into the relevant final state.
Similarly,
conversion to an expected number of $\psi (3770)$ events
required in a tau-charm factory
requires multiplication by two,
to incorporate the branching ratio of $\psi (3770)$
into two neutral $D$ mesons,
and by the inverse branching ratio for the further decays
into the relevant final states.
Interference effects in the correlated decays can be neglected
because T and CPT violation is assumed small.

We first consider the asymmetries arising from the
fixed-target experiments discussed in section III.

For the asymmetry $A_f$ given by Eq.\ \rf{if},
assuming sufficient suppression by $x$,
the number of $D^0$ events required to measure $\Re y_f$
to within one standard deviation $\si$ is
\beq
N_D (\Re y_f) \simeq
\fr 1 {8 \si^2 {\rm BR}(D^0 \rightarrow f)}
\quad .
\label{ivaa}
\eeq

The asymmetries $A_S$ and $A_L$ given in
Eqs.\ \rf{iif} and \rf{iiif} have the same general form
as $A_f$,
so the number of $D^0$ events needed to reduce the
error in either to one standard deviation
is given by an expression analogous to Eq.\ \rf{ivaa}.
However,
the interesting information is contained in
their sum and difference
$A_L \pm A_S$ given in
Eqs.\ \rf{iig} and \rf{ig}.
Combining errors in quadrature
gives the number of $D$ particles required to reduce the
error in $A_L \pm A_S$ to one standard deviation as
\beq
N_D (A_L \pm A_S) \simeq
\fr{1}{\si^2 {\rm BR}(D^0 \rightarrow \overline{K^0} + any)}
\quad .
\eeq
In particular,
for small $x$ we obtain
\beq
N_D (\Re \de_K) \simeq
N_D (\Re (\ep_K + y_K)) \simeq
\fr1{16\si^2 {\rm BR}(D^0 \rightarrow \overline{K^0} + any)}
\quad .
\label{x}
\eeq

The branching ratios relevant to the above neutral-$D$ decay
are typically of the order of several percent.
Disregarding (potentially important) experimental effects,
the present availability of $10^5$ reconstructed events
suggests that bounds of the order of $10^{-2}$ to $10^{-3}$
could be placed on both direct CPT violation in the $D$ system
(using Eq.\ \rf{ivaa})
and indirect CPT violation in the $K$ system
(using Eq.\ \rf{x}).
Direct bounds at this level would already be of interest,
and the possibility of substantially increased numbers
of reconstructed $D$ events
suggest significant improvement could be expected
in the near future.

Next,
we turn to bounds arising from experiments at
a tau-charm factory.

The parameter $\Re y_f$
determining direct CPT violation
can be measured using $A^+_f$ given in Eq.\ \rf{a}.
The second final state is unrestricted,
so it suffices to multiply by the inverse branching ratio
for the process $D^0 \rightarrow f$.
However,
the asymmetry involves only those
events for which the decay into $f$ occurs first.
Since $\ga_S\approx\ga_L$ in the $D$ system,
events with the decay into $f$ occurring first
are roughly equally frequent
as those with the decay occurring second.
An extra factor of two is therefore needed.
Collecting these factors,
we find that the number
$N_{\psi (3770)}(\Re y_f)$ of $\psi (3770)$ events
needed to reduce the error in $\Re y_f$
to within one standard deviation $\si$ is
\beq
N_{\psi (3770)}(\Re y_f) \simeq
\fr 1 {2\si^2{\rm BR}(D^0\rightarrow f)}
\quad .
\label{va}
\eeq

The combination $A^-_{L,S}$ of asymmetries given by
Eq.\ \rf{ive} measures $\Re \de_K$,
the parameter for indirect CPT violation in the $K$ system.
The independence of $A^-_{L,S}$
on the specific semileptonic-type $f$ decay
means that the corresponding branching ratios can be added.
Using
ref.\ \cite{pdt},
we get $\sum_f {\rm BR}(D^0 \rightarrow f) \simeq 27\%$.
The quantity $A^-_{L,S}$
is also independent of the specific final-state particles
that appear with the $K_S$ or $K_L$ in the other channel
in a general semileptonic-type $\overline{K^0} X$ final state.
Summing over these gives
$\sum_X {\rm BR}(D^0 \rightarrow \overline{K^0} X)
\simeq 38\%$.
Furthermore,
in determining the final result it is reasonable
to take as roughly equal the errors in the two asymmetries
$A_{f,K_L}$ and $A_{f,K_S}$.
Collecting this information,
we find that the number $N_{\psi (3770)}(\Re \de_K)$
of $\psi (3770)$ events needed to reduce the error
in $\Re \de_K$  to within one standard deviation $\si$ is
\beq
N_{\psi (3770)}(\Re \de_K) \simeq \fr{1}{\si^2}
\quad .
\eeq
This result is somewhat more favorable than the corresponding
fixed-target result,
Eq.\ \rf{x}.

The sum of the asymmetries $A^+_{L,S}$
measuring the combination $\Re (\ep_K + y_K - y_f)$,
given in Eq.\ \rf{ve},
depends on both final states.
Otherwise the estimation
is the same as for $A^-_{L,S}$ given above.
Therefore,
we get
\beq
N_{\psi (3770)}(\Re (\ep_K + y_K - y_f)) \simeq
\fr{1}{8\si^2{\rm BR}(D^0 \rightarrow \overline K^0
+ any) {\rm BR}(D^0 \rightarrow f) }
\quad .
\label{xx}
\eeq
In this case,
the result is less favorable than the corresponding
fixed-target results,
Eq.\ \rf{ivaa} and \rf{x},
because two (relatively small) branching ratios enter
instead of one.

Equations \rf{x} and \rf{xx}
involve $\ep_K$,
which measures indirect T violation in the $K$ system,
along with the parameters $y_f$ and $y_K$
for direct CPT violation in the $D$ system.
These equations can therefore be used
either as measurements of direct CPT violation
if $\ep_K$ is taken from other experiments in the $K$ system,
or as measurements providing a new check on $\ep_K$
if CPT violation is assumed small.
In the latter case,
the possibility of $10^8$ reconstructed $D$ events
could in principle produce a measurement of $\ep_K$
to $10^{-4}$ or so.\footnote{The
possibility of measuring $\ep_K$ using reconstructed
\it charged \rm $D^\pm$ events has recently
been suggested \ct{zx}.}

In the event that the magnitude of $x$ is
relatively close to the current experimental limit,
as could happen in extensions of the standard model,
it may be possible to use the above asymmetries
to extract the parameters $\Re \de_D$ and $\Im\de_D$
describing $D$-system CPT violation
and therefore to provide a test
of the string-inspired relation
\ct{kp4}
\beq
\Re\de_D = \pm \fr {2\De m}{\De\ga} \Im\de_D
\quad .
\eeq
For illustrative purposes,
we now neglect $y_f$, $x_f$, $\overline x_f$,
and terms involving second and higher powers of $x$
in the asymmetries,
and we suppose $\ep_K$ is known to sufficient precision
from other experiments.
{}From Eq.\ \rf{if} for the asymmetry $A_f$,
summing over final states $f$ we find
\beq
N_D(\Re\de_D + 2 \Im \de_D) \simeq
\fr{3}{2x^2\si^2}
\simeq \fr {600}{\si^2}
\quad ,
\eeq
where in the final form we have taken a
value $x\simeq 0.05$ close to the maximum possible.
Similarly,
summing over final states $\overline K X$ in
Eq.\ \rf{iig} gives
\beq
N_D(\Re\de_D) \simeq
\fr{1}{2x^2\si^2}
\simeq \fr {200}{\si^2}
\quad .
\eeq
Finally,
summing over final states in Eq.\ \rf{jk} gives
\beq
N_{\psi (3770)}(\Re\de_D +2 \Im\de_D ) \simeq
\fr{9}{x^2\si^2}
\simeq \fr {3600}{\si^2}
\quad .
\eeq
These equations show that both $\Re\de_D$ and
$\Im \de_D$ can in principle be extracted
if conditions are favorable.
For this purpose,
fixed-target experiments
are somewhat better from the theoretical viewpoint.
Under these circumstances,
interesting bounds could already be placed on
$\de_D$ with existing data.

\vglue 0.6cm
{\bf\noindent VI. Summary}
\vglue 0.4cm

In this paper,
the possibility of testing CPT invariance
in the neutral-$D$ system has been examined.
We give asymmetries relevant to this issue
that can be obtained from data taken at present and future
fixed-target and factory experiments.
They permit the determination of certain parameters
governing CPT violation in the $D$ and $K$ systems.
Unsuppressed measurements of direct $D$-system CPT violation
are feasible.
Moreover,
suppression by the mixing parameter $x$
of indirect $D$-system CPT violation
makes unsuppressed measurements
of indirect $K$-system CPT violation possible too.
Under particularly favorable circumstances,
indirect $D$-system CPT violation may also be measurable.

In the fixed-target case,
assuming small $x$,
Eq.\ \rf{if} gives
the parameter $\Re y_f$ controlling direct CPT violation
in the $D$ system.
Similarly,
Eq.\ \rf{ig} gives $\Re \de_K$,
involving indirect $K$-system CPT violation,
and Eq.\ \rf{iig}
gives the combination $\Re (\ep_K + y_K)$
of quantities measuring indirect $K$-system T violation
and direct $D$-system CPT violation.
All these asymmetries also contain
terms higher than second order in the small mixing parameter $x$,
involving indirect CP violation in the $D$ system.

For experiments at a tau-charm factory,
Eq.\ \rf{a} gives $\Re y_f$,
Eq.\ \rf{ive} gives $\Re \de_K$,
and Eq.\ \rf{ve} gives the combination
$\Re (\ep_K + y_K - y_f)$.
Again,
parameters for indirect CPT violation in the $D$ system
are suppressed by some power of $x$.
The only asymmetries requiring knowledge
of the sign of the time difference between the two
decays of the correlated pair
are those in Eqs.\ \rf{a} and \rf{jk}.

For both types of experiment,
excluding possible background or acceptance issues,
estimates of the bounds attainable are given in section 4.
{}From a purely theoretical perspective,
fixed-target experiments are preferable for
measuring direct CPT violation
because only one decay channel is involved.
In contrast,
experiments at a tau-charm factory
are preferable for measuring indirect $K$-system CPT violation
because the correlations between the $D$ pairs
make possible a sum over decay channels.
In any event,
interesting bounds on various types of CPT violation
are attainable in the neutral-$D$ system,
using data already existing
or likely to become available within a few years.

\vglue 0.4cm
{\it\noindent Acknowledgment.}
This work was supported in part
by the United States Department of Energy
under grant number DE-FG02-91ER40661.

\baselineskip=18pt

\vglue 0.4cm

\end{document}